\documentclass{aa}  

\usepackage{graphicx}
\usepackage{txfonts}
\usepackage{lipsum}
\usepackage{subcaption}         

\usepackage{lscape}            
\usepackage{placeins}          
\usepackage{xcolor}
\usepackage{stfloats}
\usepackage[pdfpagelabels=false]{hyperref} 
\hypersetup{colorlinks=true,linkcolor=black,citecolor=blue,filecolor=blue,urlcolor=blue}

\begin{document}

   \title{The influence of hypothetical exomoons on planetary thermal phase curves}

   \author{Xinyi Song\inst{1}\,$^{\href{https://orcid.org/0000-0003-4972-7772}{\protect\includegraphics[height=0.19cm]{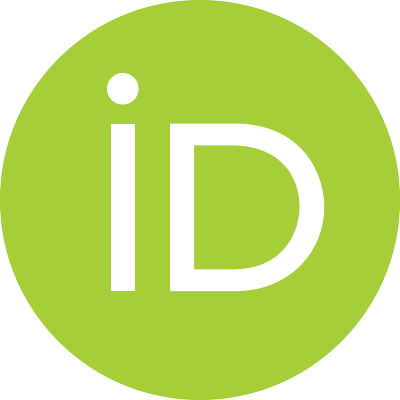}}}$,
           Jun Yang\inst{1}\thanks{Corresponding author: \texttt{junyang@pku.edu.cn}},
           \and Yueyun Ouyang\inst{1}
           }
           
    \institute{Laboratory for Climate and Ocean-Atmosphere Studies, Department of Atmospheric and Oceanic Sciences, \\School of Physics, Peking University, Beijing 100871, China
    }
        
   \date{Received XXX; accepted XXX}

  \abstract
   {More than 200 moons exist in our Solar System, yet no exomoon has been confirmed to date. While the innermost two planets of the Solar System lack natural satellites and most studies favour the existence of exomoons around long-period planets, some theoretical studies that take tidal dissipation, orbital decay, and migration processes into account suggest that exomoons may survive around short-period exoplanets.}
   {We investigated the impact of exomoons on planetary thermal phase curves and assessed their detectability within a theoretical framework.} 
   {We simulated the thermal phase curves of exomoon–exoplanet systems, including mutual transits and occultations, and explored their dependence on planetary orbital periods across a wide range of systems.}
   {Close-in airless exomoons maintain large day–night temperature contrasts, amplifying the thermal phase-curve signal of the system. When the exomoon transits or is occulted by the exoplanet, the transit depth varies with the planetary phase, and the occultation depth varies with the exomoon's phase. The maximum occultation depth can reach $\sim 20$ ppm for long-period systems. For short-period planets, the signal can reach up to $\sim$100 ppm, although such configurations may not be dynamically stable over long timescales.}
   {If exomoons are not accounted for, the planetary temperature distribution retrieved from observed thermal phase curves may overestimate the planetary day--night temperature contrast and underestimate the planetary horizontal heat transport. In principle, the periodic exomoon–exoplanet mutual occultation signal could be extracted using methods such as box-fitting least squares, providing a framework for future observational studies and instrument planning.}


   \maketitle

\section{Introduction}
Since the discovery of 51 Pegasi b in 1995 \citep{Mayor_1995}, more than 6000 exoplanets have been discovered. However, their satellites, or exomoons, remain undetected. Our Solar System has more than 200 moons orbiting six planets, and these moons provide valuable information on planetary formation, bombardment, dynamics, and planetary system evolution \citep{Heller_2014,Barr_2016,heller2018detecting}. The existence of the Moon is one of the main reasons why Earth’s obliquity variations are small, and a stable obliquity may be crucial for land life to develop \citep{Laskar_1993,kasting2001peter}.
The icy moons in our Solar System have liquid water under icy surfaces \citep{Squyres_1983,Kivelson_2002,Kuskov_2005,Vance_2014}. The potential habitability of these icy moons is one of the main interests of several current and near-future space missions \citep{Grasset_2013,Phillips_2014,Jackson_2020}. Beyond our Solar System, tidal heating and incident planetary radiation may make exomoons potential environments for habitability. \citep{Heller_2013,Heller_2014}. With ever-improving space telescopes, we are on the brink of making the first exomoon detection. To better propose or exclude more exomoon candidates, or even confirm an exomoon candidate, both detection and theoretical studies of exomoons are worth researching. 

Dozens of approaches for detecting exomoons have been proposed, including direct imaging, microlensing, astrometry, radial velocity (RV), and transit methods. The RV method measures velocity variations induced by gravitational interactions. For exomoons, however, the gravitational influence on the host star's RV signal is far too weak to be detected \citep{teachey2024detecting}. As a result, exomoons are usually not detectable through conventional stellar RV measurements. Detecting exomoons via the RV method therefore requires directly imaging the host planets, and such measurements are typically carried out in the H band ($\sim$1.65 $\mu$m) or K band ($\sim$2.2 $\mu$m; \citealt{Ruffio_2021}). However, the existing telescopes can only achieve sub kilometre-per-second precision in these bands \citep{Ruffio_2021}, which is  insufficient to detect exomoons. With such a precision level, we can only rule out exomoons larger than a half Jupiter mass \citep{Vanderburg_2021}.

Searching for exomoons through transits is the most widely used strategy \citep{Kipping_2012,Heller_2024} since most of the known exoplanets are transiting planets.
Several studies have been dedicated to finding transiting exomoons by analysing transit timing variations \citep{Simon_2007,Kipping_2009}, transit duration variations \citep{Kipping_2009}, or transit depth variations \citep{Rodenbeck_2020}. Kepler-1625 b and Kepler-1708 b, nearly Neptune-sized potential exomoons around Jupiter-sized exoplanets, are two of the most promising transiting candidates \citep{Teachey_2018, Kipping_2022}. However, \cite{Heller_2024} find that the claimed exomoon transit signals may be caused by stellar activity, limb-darkening, or data fitting processes. Other candidates, such as one candidate around a directly imaged brown dwarf, DH Tauri B \citep{Lazzoni_2020}, and another around an isolated planetary-mass object \citep{Limbach_2021}, remain unconfirmed.

The thermal phase curve can also be used to detect exomoons. The thermal phase curve shows how globally integrated thermal emission flux varies along the orbital phase. The variations in the thermal phase curve contain information about the planetary temperature distribution.
In a planet--moon system, both the planet and the moon contribute to the variation in the thermal phase curves. In our Solar System, the Moon is airless and slow-rotating, so the dayside temperature can reach 400 K, while the nightside temperature is just about 100 K. On Earth, the day--night temperature difference is just about 10 K \citep{Pithan_2024}. Although the Moon's radius is less than one-third of Earth's, the lunar day--night temperature difference is so large that the variation in the total thermal phase curve (i.e. planet + moon) is mainly caused by the Moon \citep{Moskovitz_2009,Robinson_2011,G_mez_Leal_2012}. If an exomoon is airless and large enough, the thermal phase curve variation contributed by the airless exomoon can be larger than that of the planet \citep{Gaidos_2005}. 

Early studies of exomoon detection using thermal phase curves mainly focused on disentangling planetary and lunar signals through the amplitude, phase, or spectral contrast of the combined emission. For example, \citet{Moskovitz_2009} showed that the phase-curve amplitude and phase alone are insufficient to separate planetary and lunar contributions for Earth–Moon analogues, while \citet{Forgan_2017} found that even spectral discrimination remains challenging unless the exomoon is strongly tidally heated or intrinsically luminous. These results highlight the limitations of phase-curve-only approaches when applied to temperate, Earth-like systems.

Tidally heated exomoons can be much brighter than their host exoplanets, and some can be directly imaged \citep{Limbach_2013}. For close-in, volcanically active exomoons, also known as exo-Ios, the outgassing can also be detected via the phase variability in the thermal phase curves \citep{Oza_2019,Kleisioti_2024,Meyer_zu_Westram_2024}. A recent study by \cite{Oza_2024} found a possible transient signal of an exo‐Io.

For non-tidally heated exomoons or planetary companions, detection via the thermal phase curve is possible through mutual events, such as transits and shadows \citep{Cabrera_2007,Limbach_2021}. 
\cite{Limbach_2024} analysed Earth–Moon mutual event observations and showed that Earth–Moon analogues within 10 pc of the Sun are detectable with $\sim$2--20 mutual events. 
However, exoplanetary systems can be very different from the Jupiter--Io system or the Earth--Moon system. The day--night temperature difference, orbital and rotation periods, masses, and radii of the moon and the planet all influence the thermal phase curves to varying degrees. A wider parameter space and more calculations are necessary to verify whether the method of extracting exomoon signals based on the Earth--Moon system is appropriate for exoplanetary systems. 
In this work we focus on the influence of non-tidally heated exomoons on planetary thermal phase curves and discuss the possibility of detecting exomoons through mutual transits and eclipses on a broader scope.

From an observational perspective, airless exomoons orbiting close-in exoplanets are more likely to have high day--night temperature contrasts. This observational advantage, however, is counterbalanced by dynamical considerations: several studies have suggested that the existence of exomoons around close-in exoplanets (planets with small orbital distances or short orbital periods) is dynamically unfavourable. During the post-formation tidal evolution stage, an exomoon orbiting faster than the planetary rotation tends to spiral inwards towards the Roche limit, and an exomoon orbiting more slowly evolves outwards and may escape the Hill sphere \citep{Counselman_1973,murray1999solar}. Based on tidal migration theory, \cite{Barnes_2002} argued that exomoons around close-in gas giants are likely to have very short lifetimes, and other studies considering planetary and lunar tidal interactions found that exomoon stability increases significantly for planets on longer orbital periods \citep{Sasaki_2012,Sucerquia_2019,Dobos_2021}.

However, subsequent studies found that exomoons surviving around close-in planets is definitely not impossible. Using a finer derivation of the planet--star mass ratio, \cite{Domingos_2006} demonstrated that exomoons within $\sim$0.49 Hill radii remain stable, and that even Earth-like exomoons can be stable around exoplanets in the habitable zone. \cite{Cassidy_2009} pointed out that for gas giants such as Jupiter, the appropriate value of the planetary tidal dissipation factor ($Q_p$) may be much higher than that used by \cite{Barnes_2002}, by a factor of $\sim$10$^8$. With the much larger $Q_p$ and taking into account the dynamical changes caused by the lunar tidal response, \cite{Cassidy_2009} argued that even Earth-size exomoons around hot Jupiters can survive the age of the Solar System. A more recent study shows that tidal dissipation in exomoons can prevent  escape from the planet’s Hill sphere, keeping the exomoons bound to their planets at a semi-major axis of $\sim$0.4 Hill radii \citep{Kisare_2023}, which means that exomoons may be more stable than previously thought.

The migration history of the planet in the protoplanetary disk also influences the orbital stability of exomoons. Although some studies showed that exomoons are likely to be ejected or collide with the planet during migration \citep{Namouni_2010,spalding2016resonant}, exomoons captured from the protoplanetary disk on retrograde orbits can survive around close-in exoplanets \citep{Namouni_2010}.

If an exomoon exists around a close-in exoplanet, how much can such an exomoon possibly influence the total thermal phase curve? In Sects.~\ref{TRAPPIST} and~\ref{GJ} we calculate the thermal phase curves of two close-in exomoons to explore the upper limits of possible exomoon signals. Apart from the  radius and day--night temperature contrast, what other factors can influence an exomoon's thermal phase curves? In Sect.~\ref{subsec:cp} we explore the influence of the surface thermodynamic parameters and the rotation period of the exomoon on thermal phase curves. An extension to exomoons orbiting longer-period planets is presented in Sect.~\ref{subsec:long_period}. Section~\ref{sec:discussions} provides a summary and discussions.

\section{Methods}
\subsection{Exomoon orbital parameters} \label{subsec:orbit}
For a wide range of exoplanets, we calculated the thermal phase curves of their possible exomoons. As a first step, we determined the orbital parameters of the exomoons.

We assumed the eccentricity to be zero. The semi-major axis between the exomoon and the planet was set to 0.4 Hill radii, the outer boundary in \cite{Kisare_2023}, so that the exomoon could maintain the largest day--night temperature contrast.

The Hill radius describes the possible outermost orbit of the moon. A moon can remain gravitationally bound to the planet only within the Hill radius. In our Solar System, all known natural satellites lie well within their planet’s Hill sphere. The Hill radius is given by
\begin{equation}
R_{Hill} = a_{planet} \sqrt[3]{\frac{M_{planet}}{3(M_{star}+M_{planet})}},
\label{equation_hill}
\end{equation}
where $a_{planet}$ is the planetary semi-major axis, and $M_{star}$ is the mass of the star \citep{lissauer2013fundamental}.

We chose four planets for detailed case studies, spanning close-in and longer-period regimes: TRAPPIST-1 e, a terrestrial planet in the habitable zone; GJ 1214 b, a close-in sub-Neptune; and two relatively long-period planets, TOI-2525 b ($P=23.3$ days) and CoRoT-9 b ($P=95.3$ days).
For more details on these four systems, see Table~\ref{table_orbit}. 

\begin{table*}[ht!]
    \caption {Orbital information of four hypothetical exomoon systems. }
    \label{table_orbit} 
    \def\arraystretch{1.2}
    \centering
    \begin{tabular}{ccccc}
    \hline\hline 
    Host star & Host planet & Planetary orbit (Star-centric) & Lunar orbit (Planetary-centric) & Average Flux\\
    \hline
    TRAPPIST-1 & TRAPPIST-1 e & $ $ & $ $ &  \\
    $M_s$ = 0.08 $M_{Sun}$ & $M_p$ = 0.69 $M_{e}$ & $a_{p}$ = 0.02817 AU & $a_{m}$ = 33300 km (5.68 $R_p$) & 224.9 W$m^{2}$\\
    $R_s$ = 0.117 $R_{Sun}$ & $R_p$ = 0.92 $R_{e}$ & Period = 6.10 day & Period = 0.876 day \\
    \hline
    GJ 1214 & GJ 1214 b & $ $ & $ $ &  \\
    $M_s$ = 0.15 $M_{Sun}$ & $M_p$ = 8.41 $M_{e}$ & $a_{p}$ = 0.01411 AU & $a_{m}$ = 32313 km (1.85 $R_p$) & 6027.6 W$m^{2}$\\
    $R_s$ = 0.216 $R_{Sun}$ & $R_p$ = 2.74 $R_{e}$ & Period = 1.58 day & Period = 0.231 day \\
    \hline
    TOI-2525 & TOI-2525 b & $ $ & $ $ &  \\
    $M_s$ = 0.849 $M_{Sun}$ & $M_p$ = 27 $M_{e}$ & $a_{p}$ = 0.1511 AU & $a_{m}$ = 285339 km (5.16 $R_p$) & 5437.0 W$m^{2}$\\
    $R_s$ = 0.785 $R_{Sun}$ & $R_p$ = 8.68 $R_{e}$ & Period = 23.28 day & Period = 3.398 day \\
    \hline
    CoRoT-9 & CoRoT-9 b & $ $ & $ $ &  \\
    $M_s$ = 0.993 $M_{Sun}$ & $M_p$ = 267 $M_{e}$ & $a_{p}$ = 0.4021 AU & $a_{m}$ = 1570880 km (20.63 $R_p$) & 1757.7 W$m^{2}$\\
    $R_s$ = 0.898 $R_{Sun}$ & $R_p$ = 11.95 $R_{e}$ & Period = 95.27 day & Period = 13.880 day \\
    \hline
    \multicolumn{5}{l}{$M_{e}$ and $R_{e}$: the mass and radius of Earth. $M_{J}$ and $R_{J}$: the mass and radius of Jupiter.}\\
    \end{tabular}
\end{table*}

\subsection{The mass and radius of the exomoon} \label{subsec:mass-radius}
Our study focuses on the influence of large exomoons on the thermal phase curve. For each case, we studied exomoons with three different hypothetical sizes, Moon size (1738 km) and 5\% and 10\% of the planetary mass, in order to probe the largest possible exomoon thermal phase curve signal.
For moons with radius below 1.5 Earth radii, the density-radius relation is given by
\begin{equation}
\rho_{moon} = 2.43 + 3.39(R_{moon}/R_{e}),
\label{equation_density}
\end{equation}
where $\rho_{moon}$ is the density of moon in units of g cm$^{-3}$, $R_{e}$ is Earth's radius, 6371 km \citep{Weiss_2014}. Knowing the mass of the moon, with $M_{moon}=\frac{4}{3}\pi \rho_{moon}R_{moon}^3$, we calculated the density and radius, and calculated the corresponding Hill radius.

\subsection{Exomoon surface temperature calculation} \label{subsec:SST}
We considered two types of exomoons: exomoons with thick atmosphere and airless exomoons.
For exomoons with thick atmosphere, assuming the heat transport efficiency is extremely high, the temperature is uniform around the globe, and the emissivity is unity. The effective emission temperature ($T_{eff}$) is given by 
\begin{equation}
T_{eff} = \left[\frac{(1-\alpha)L_{star}}{16\sigma\pi a^2}\right]^{1/4},
\label{equation_Teff}
\end{equation}
where $L_{star}$ is the luminosity of the host star, $\sigma$ is the Stefan–Boltzmann constant, and $\alpha$ is the bond albedo of the exomoon, which is set to 0.3.

For airless exomoons, we used the code developed by \cite{ouyang2024potential} to calculate the surface temperature distribution of exomoons. Internal heat source was not included in our simulations. Because the vertical temperature gradients are
typically much greater than the horizontal gradients in the subsurface layers, the code only considers vertical heat conduction, and ignores horizontal heat advection. The energy equation is 

\begin{equation}
\rho_s c_{ps}\partial T / \partial t = \partial/\partial_z\left(\kappa_T \partial T / \partial z\right),
\label{equation_Tsoil}
\end{equation}
where $\rho_s$, $c_{ps}$, and $\kappa_T$ are the density, specific heat capacity, and thermal conductivity of the subsurface layer, respectively, and $z$ is the vertical coordinate.

For the surface boundary condition in the airless scenario, the diffusive heat flux at the surface equals the net flux between insolation heating and thermal emission:

\begin{equation}
    \kappa_T\partial_z T|_{z=0} = \left(1-\alpha\right)S(t)-F(T_s),
    \label{equation_Tboundary}
\end{equation}
where $S(t)$ is the incident stellar flux, $F(T_s) = \sigma T_s^4$ is the infrared emission, and $T_s$ is the surface temperature. The model assumed a surface infrared emissivity of 1. 

The subsurface soil consisted of 35 layers (Eq. 8 in \citealt{ouyang2024potential}), and the deepest layer was in scale of the characteristic depth $Z=\sqrt{\kappa_T \Delta t/\rho_s c_{ps}}$. The characteristic depth ($Z$) was the deepest depth to which temperature fluctuation could reach in a single time step ($\Delta t$). For all of our simulations, the deepest subsurface soil layers were within 2 m. The surface albedo was 0.136 (uniform) over the exomoon surface. The model had 37 grids over the latitude and 73 grids over the longitude. The time step was 3600 seconds for the TRAPPIST-1 e system and 180 seconds for the GJ 1214 b system. The total integration time was 50 planetary orbital periods for the TRAPPIST-1 e system and 10 planetary orbital periods for the GJ 1214 b system, so that the surface temperature distribution pattern reaches stability for at least the last 50\% of integration time. We used snapshots from the final simulated year to calculate the thermal phase curves, calculating the ratio of the planetary or lunar infrared emission to the stellar infrared emission.

\section{Results\label{results}} 

\subsection{TRAPPIST-1 e with a hypothetical exomoon\label{TRAPPIST}}

\begin{figure*}[ht!]
    \includegraphics[width=0.95\linewidth]{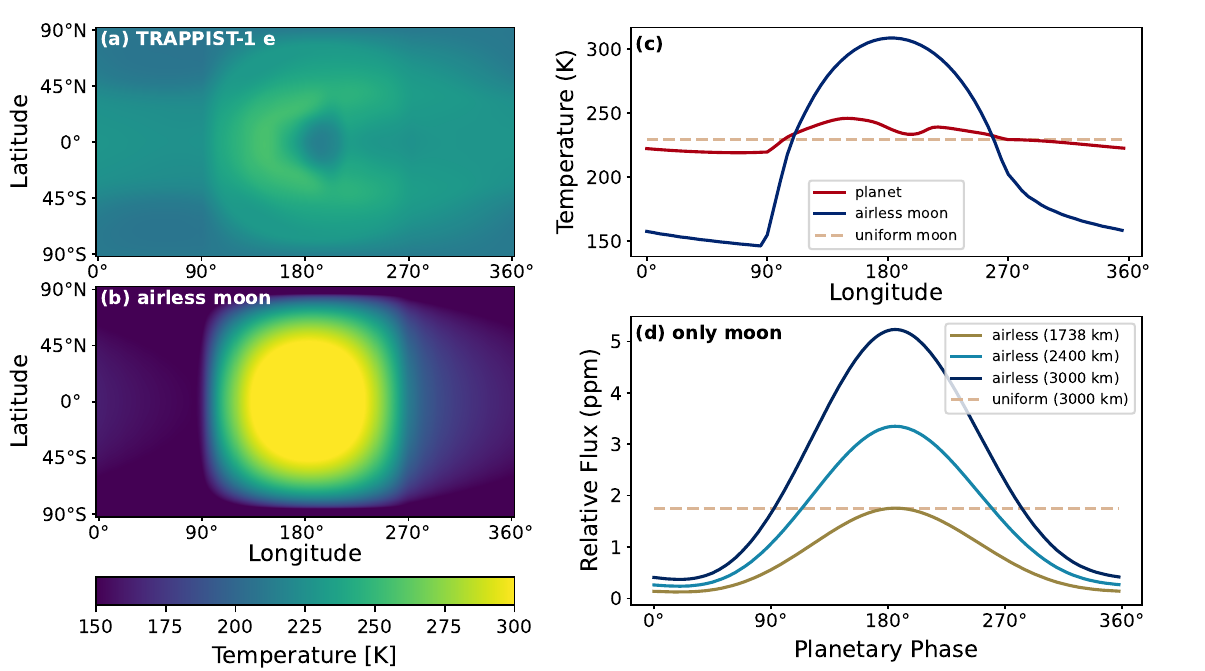}
    \centering
    \caption{Temperature distributions and thermal phase curves of the TRAPPIST-1 e system. Panel (a): Snapshot of the effective emission temperature of TRAPPIST-1 e. Panel (b): Snapshot of the surface temperature distribution of a hypothetical tidally locked airless exomoon. Panel (c): Meridional mean effective emission temperatures of: TRAPPIST-1 e (red line), a tidally locked airless exomoon (blue line), and an exomoon with uniform temperature (dashed line). Panel (d): Exomoon thermal phase curves, i.e. the ratio of the lunar infrared emission to the stellar infrared emission. The green, cyan, and dark blue lines represent tidally locked airless exomoons with radii of 1738 km, 2400 km, and 3000 km, respectively. The dashed line shows the thermal phase curve of an exomoon with uniform temperature and a radius of 3000 km. The exomoons orbit at 33300 km around TRAPPIST-1 e, and the orbital period is 0.876 Earth days. The thermal phase curves are integrated over wavelengths from 5 to 50 $\mu$m.}
    \label{fig:trappist-upper}
\end{figure*}

Figure~\ref{fig:trappist-upper}a shows a snapshot of the effective emission temperature of TRAPPIST-1 e. The day--night temperature contrast is relatively low, just about 40 K, due to horizontal atmospheric heat transport. With the effect of super-rotation, high clouds around the substellar point (180$^{\circ}$) shift slightly eastwards, causing the meridional mean temperature around 190$^{\circ}$ to be slightly cooler than the temperature of surrounding areas. The data are from global circulation model (ExoCAM) simulations in \citet{Yang_2023}. The simulation is for an aqua-Earth, with surface gravity 0.93 of Earth's value, and the mean surface air pressure is $\sim$0.93 bar (N$_2$) with a CO$_2$ concentration of 355 ppmv. We used this simulation result to test the influence of hypothetical exomoons on the thermal phase curve of a habitable planetary system around a red dwarf.

Figure~\ref{fig:trappist-upper}b shows a snapshot of the surface temperature of the hypothetical airless exomoon. The substellar point is 180$^{\circ}$. The hottest surface temperature reaches above 300 K and the lowest temperature reaches below 150 K. As suggested by Eq.~\ref{equation_Teff}, if the moon reaches thermal equilibrium instantly, the substellar temperature $T_{sub} = \left[\frac{(1-\alpha)L_{star}}{\sigma4\pi a^2}\right]^{1/4} = 342$ K, which is $4^{0.25}$ times the global-mean effective emission temperature. In realistic situations, because of the thermal inertia of the surface, heating the substellar point takes time, so the substellar temperature is cooler than the instant equilibrium.

Figure~\ref{fig:trappist-upper}c shows the meridional (north--south) mean effective emission temperature of TRAPPIST-1 e and the hypothetical exomoons.
For a tidally locked airless exomoon orbiting at 33300 km around TRAPPIST-1 e, both the orbital and rotation periods are 0.876 Earth days. Despite the short rotation period, the day--night temperature difference of the exomoon can still reach more than 150 K (blue line in Fig.~\ref{fig:trappist-upper}c) because of the low surface heat capacity and thermal conductivity. In the surface temperature simulations of exomoons around TRAPPIST-1 e, the surface has the same heat capacity ($\rho c_p = 10^{6}$ J m$^{-3}$ K$^{-1}$) and thermal conductivity ($\kappa_T = 0.01$ W m$^{-1}$ K$^{-1}$) as the regolith of the Moon \cite[Chapter~7.4]{pierrehumbert2010principles}. 

For an exomoon with a thick atmosphere, assuming that the temperature distribution is uniform and the bond albedo is 0.3, the effective emission temperature is 230 K (dashed line in Fig.~\ref{fig:trappist-upper}c). The red line in Fig.~\ref{fig:trappist-upper}c shows the meridional mean effective emission temperature of TRAPPIST-1 e.

\begin{figure*}[ht!]
    \includegraphics[width=0.95\linewidth]{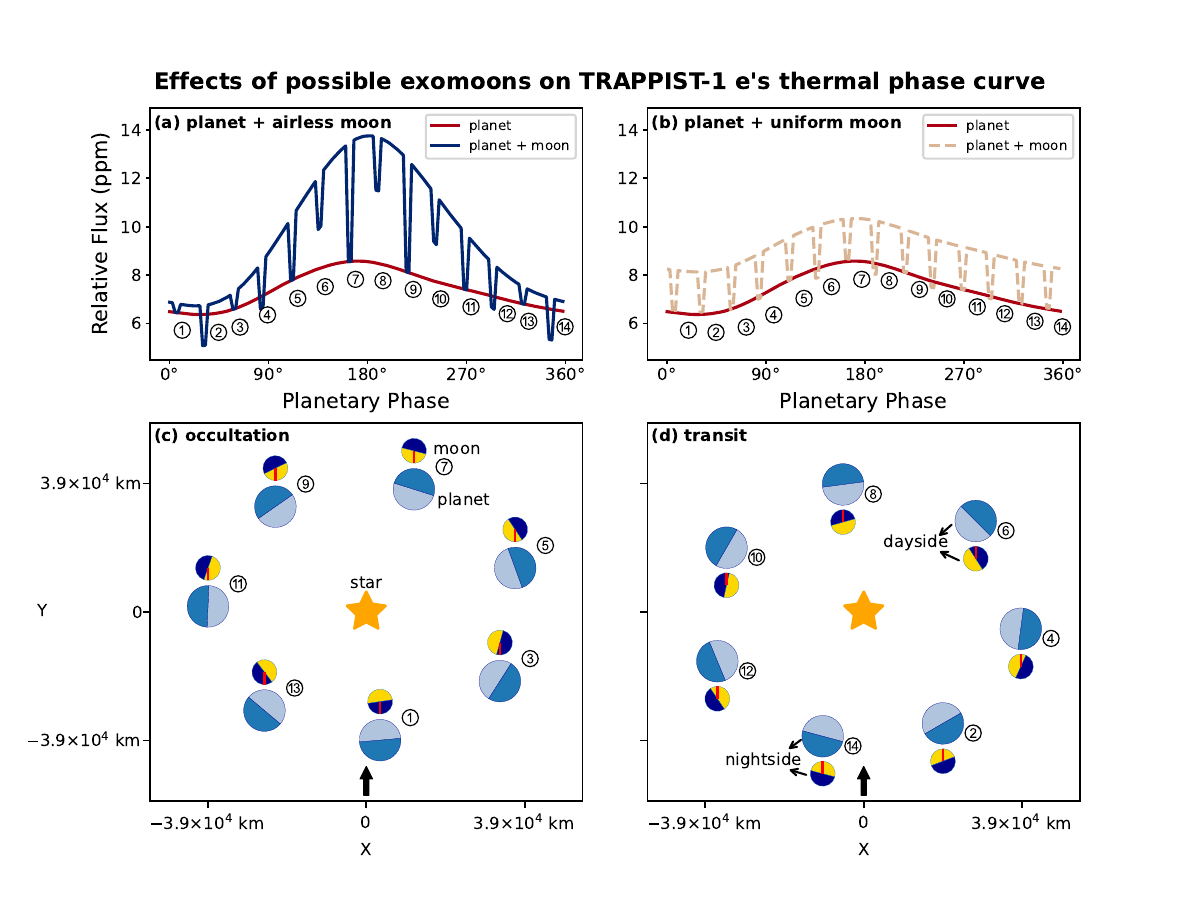}
    \centering
    \caption{Thermal phase curves and eclipses for the TRAPPIST-1 e system. Panel (a): Total thermal phase curve of the planet and a 3000 km tidally locked airless exomoon (blue line). Panel (b): Total thermal phase curve of the planet and a 3000 km exomoon with uniform temperature (dashed line). The exomoons orbit at 33300 km around TRAPPIST-1 e. The red lines in panels (a) and (b) show the planetary phase curve alone, i.e. the ratio of the planetary infrared emission to the stellar infrared emission. Panel (c): Occultation scenarios, with the exomoon blocked by the planet. The red indicators on the exomoon markers show the 180$^{\circ}$ longitude. Panel (d): Transit scenarios, with the exoplanet blocked by the exomoon. The observation direction in panels (c) and (d) is from bottom to top, as indicated by the black arrows at the bottom.}
    \label{fig:trappist-lower}
\end{figure*}

Three factors influence the thermal phase curve of an exomoon-exoplanet system: thermal flux from the exomoon, thermal flux from the exoplanet, and infrared emission from the host star. To calculate the thermal flux of the exomoon, we assumed that the exomoon is a blackbody and integrate the thermal flux over 5--50 $\mu$m, to be consistent with the planetary outgoing thermal flux of the ExoCAM data. To calculate the stellar thermal flux, we used the stellar spectra from BT-Settl (AGSS2009) models \citep{allard2012models}, and integrated from 5 $\mu$m to 50 $\mu$m.

Figure~\ref{fig:trappist-upper}d shows the thermal phase curves of exomoons with different radii. Fixing the surface temperature distribution, the thermal flux observed from a celestial body is proportional to the square of its radius. If the exomoon has the same radius as the Moon, 1738 km, the peak-to-peak amplitude of the thermal phase curve is about 2 ppm. Using Eq.~\ref{equation_density}, we derive that if the exomoon has 5\% or 10\% of the planetary mass, the exomoon radius is about 2400 km (or 3000 km), and the peak-to-peak amplitude of the thermal phase curve is about 3.5 ppm (or 5.5 ppm). 

The dashed line in Fig.~\ref{fig:trappist-upper}d shows the thermal phase curve of a 3000-km exomoon with uniform temperature. When the effective emission temperature does not vary along the globe, the thermal phase curve does not vary along the phase angle, so the thermal phase curve amplitude is zero for an exomoon with uniform temperature.

The red line in Fig.~\ref{fig:trappist-lower}a shows the planetary contribution of the thermal phase curve, and its peak-to-peak amplitude is about 2 ppm. The radius of TRAPPIST-1 e is almost 5900 km, more than 3 times the Moon's radius, but the planetary day--night temperature contrast is very low, less than 40 K (red line in Fig.~\ref{fig:trappist-upper}c), so the planetary thermal phase curve amplitude (red line in Fig.~\ref{fig:trappist-lower}a) is comparable to the thermal phase curve amplitude of an airless exomoon with the Moon's radius (olive-coloured line in Fig.~\ref{fig:trappist-upper}d). 

The existence of an airless exomoon significantly amplifies the total thermal phase curve amplitude. As shown by the blue line in Fig.~\ref{fig:trappist-lower}a, an airless exomoon with radius of 3000 km almost triples the total thermal phase curve amplitude. An exomoon with uniform temperature distribution does not influence the amplitude. As shown by the dashed line in Fig.~\ref{fig:trappist-lower}b, a 3000-km exomoon with uniform temperature distribution only uniformly adds 2 ppm to the thermal phase curve signal.

The labelled dips in Fig.~\ref{fig:trappist-lower}a show the mutual transits between the hypothetical exomoon and the planet. The orbital period of TRAPPIST-1 e is 6 Earth days, and the exomoon's orbital period is 0.876 Earth days. The exomoon transits the planet seven times over a full planetary orbit. Each exomoon-exoplanet transit lasts more than 60 minutes, comparable with the 54-minute exoplanet-star transit.

When the exomoon passes behind the exoplanet, the scenario is defined as occultation, as shown in Fig.~\ref{fig:trappist-lower}c. The direction of observation is from bottom to top, as shown by the black arrow at the bottom of Figs.~\ref{fig:trappist-lower}c and \ref{fig:trappist-lower}d. During the occultation, flux from the exomoon is blocked by the exoplanet, so only the planetary signal contributes to the total thermal phase curve. The total thermal phase curves (solid blue line and dashed pink line in Figs.~\ref{fig:trappist-lower}a and \ref{fig:trappist-lower}b) coincide with the planetary thermal phase curves (red lines in Figs.~\ref{fig:trappist-lower}a and \ref{fig:trappist-lower}b), as labelled by odd numbers in Fig.~\ref{fig:trappist-lower}a.

When the exomoon passes in front of the exoplanet, the scenario is defined as a transit, as shown in Fig.~\ref{fig:trappist-lower}d. During transit, $F_{total}=\left[1-\left(\frac{R_{moon}}{R_{planet}}\right)^2\right]F_{planet}+F_{moon}$. If the nightside of the exoplanet is facing the observer during the transit, the nightside of the exomoon is much colder than the planetary area blocked by the moon, $F_{moon}<
\left(\frac{R_{moon}}{R_{planet}}\right)^2 F_{planet}$, so the total thermal flux is weaker than the planetary thermal flux, such as the dips labelled {\textcircled{{\scalebox{0.8}{2}}}}, {\textcircled{{\scalebox{0.8}{4}}}}, {\textcircled{\raisebox{.2pt} {\scalebox{0.65}{12}}}}, and {\textcircled{\raisebox{.2pt} {\scalebox{0.65}{14}}}} in Fig.~\ref{fig:trappist-lower}a.
If the dayside of the exoplanet is facing the observer during the transit, the dayside of the exomoon is much hotter than the planetary area blocked by the moon, $F_{moon}>
\left(\frac{R_{moon}}{R_{planet}}\right)^2 F_{planet}$, so the total thermal flux is greater than the planetary thermal flux, such as the dips labelled {\textcircled{\raisebox{-.5pt}{\scalebox{0.8}{6}}}, {\textcircled{\raisebox{-.5pt}{\scalebox{0.8}{8}}}, and {\textcircled{\raisebox{.2pt} {\scalebox{0.65}{10}}}} in Fig.~\ref{fig:trappist-lower}a.

\subsection{GJ 1214 b with a hypothetical exomoon \label{GJ}}
Figure~\ref{fig:GJ-upper}a shows a snapshot of the surface temperature distribution of a hypothetical exomoon around GJ 1214 b. The hottest surface temperature reaches around 700 K, and the lowest temperature is about 200 K. The stellar luminosity of GJ 1214 is very high, about 7 times of TRAPPIST-1's, so the day--night temperature contrasts of both the exoplanet and the exomoon are much higher in the GJ 1214 b system, as shown in  Fig.~\ref{fig:GJ-upper}b. The red line shows the equatorial 5--12 $\mu$m effective emission temperature of GJ 1214 b shown in \cite{Kempton_2023}. For a tidally locked airless exomoon orbiting at 32313 km around GJ 1214 b, both the orbital and rotation periods are 0.231 Earth days. The day--night temperature difference of the airless exomoon increases to over 500 K (blue line in Fig.~\ref{fig:GJ-upper}b).  For an exomoon with thick atmosphere, assuming the temperature distribution is uniform and the bond albedo is 0.3, the effective emission temperature reaches 522 K (dashed line in Fig.~\ref{fig:GJ-upper}b). The surface thermal parameters are the same as in Sect.~\ref{TRAPPIST} ($\rho c_p = 10^{6}$ J m$^{-3}$ K$^{-1}$, and $\kappa_T = 0.01$ W m$^{-1}$ K$^{-1}$). For a more quantitative discussion on the day--night temperature difference of an airless exomoon, please see Sect.~\ref{subsec:cp}.

\begin{figure}
    \includegraphics[width=\linewidth]{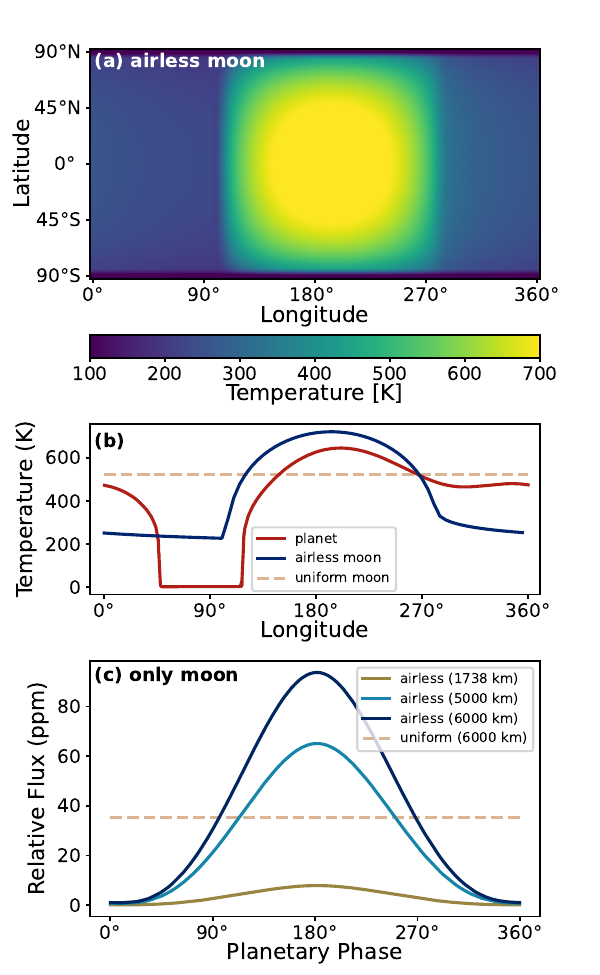}
    \centering
    \caption{Surface temperature and thermal phase curves of the GJ 1214 b system. Panel (a): Snapshot of the surface temperature distribution of a hypothetical tidally locked airless exomoon. Panel (b): Meridional mean effective emission temperatures of: GJ 1214 b (red line), a tidally locked airless exomoon (blue line), and an exomoon with uniform temperature (dashed line). The GJ 1214 b data are from \cite{Kempton_2023}. Panel (c): Exomoon thermal phase curves. The green, cyan, and dark blue lines represent tidally locked airless exomoons with radii of 1738 km, 5000 km, and 6000 km, respectively. The dashed line shows the thermal phase curve of an exomoon with uniform temperature and a radius of 6000 km. The exomoons orbit at 32313 km around GJ 1214 b, and the orbital period is 0.231 Earth days.}
    \label{fig:GJ-upper}
\end{figure}

\begin{figure*}
    \sidecaption
    \includegraphics[width=12cm]{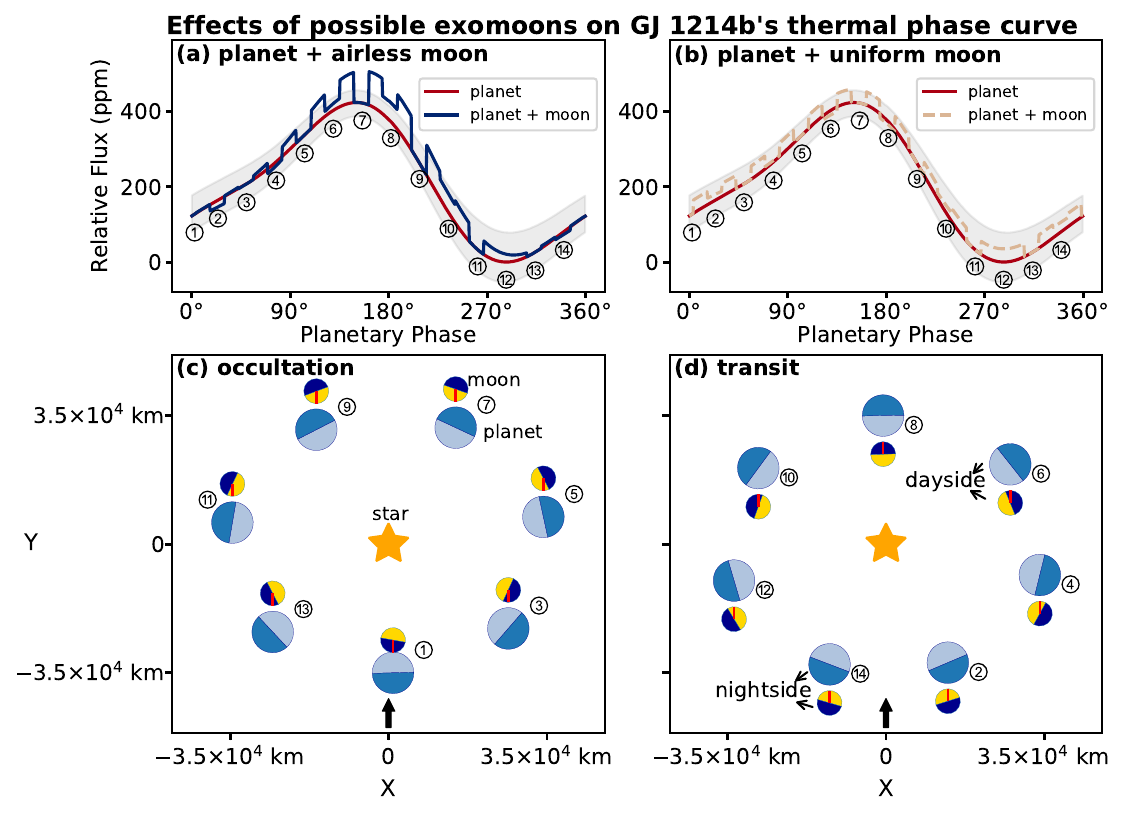}
    \caption{Thermal phase curves and eclipses for the GJ 1214 b system. Panel (a): Total thermal phase curve of the planet and a 6000 km tidally locked airless exomoon (blue line). Panel (b): Total thermal phase curve of the planet and a 6000 km exomoon with uniform temperature (dashed line). The exomoons orbit at 32313 km around GJ 1214 b. The red lines in panels (a) and (b) show the thermal phase curve of GJ 1214 b alone, with data from \cite{Kempton_2023}. Panel (c): Occultation scenarios, with the exomoon blocked by the planet. The red indicators on the exomoon markers show the 180$^{\circ}$ longitude. Panel (d): Transit scenarios, with the exoplanet blocked by the exomoon. The observation direction in panels (c) and (d) is from bottom to top, as indicated by the black arrows at the bottom.}
    \label{fig:GJ-lower}
\end{figure*}

GJ 1214 b is a sub-Neptune, so stable exomoons around it can be much larger than the exomoons around TRAPPIST-1 e. If the exomoon has 5\% (or 10\%) of the planetary mass, the exomoon's radius is about 5000 km (or 6000 km; Eq.~\ref{equation_density}). To calculate the thermal phase curve, we assumed the exomoon is a blackbody and integrated the thermal flux over 5--12 $\mu$m, in order to be consistent with the GJ 1214 b thermal phase curve observed with the \textit{James Webb} Space Telescope (JWST) Mid-Infrared Instrument Low Resolution Spectrometer \citep{Kempton_2023}. 

With radius and day--night temperature difference much larger than those of the hypothetical exomoon around TRAPPIST-1 e, the thermal phase curve peak-to-peak amplitude can reach $\sim$100 ppm for a 6000-km airless exomoon around GJ 1214 b (blue line in Fig.~\ref{fig:GJ-upper}c). The 100 ppm signal contributes about 20\% to the total thermal phase curve amplitude (blue line in Fig.~\ref{fig:GJ-lower}a). The peak-to-peak amplitude is about 400 ppm for GJ 1214 b's thermal phase curve contribution (red line in Fig.~\ref{fig:GJ-lower}a), four times the exomoon contribution, because the planetary radius is very large, 17081 km, 2.8 times of the largest exomoon radius we considered (6000 km), and also because the day--night temperature contrast on the planet is about the same order of magnitude as the contrast on the airless exomoon (red and blue lines in Fig.~\ref{fig:GJ-upper}b).

GJ 1214 b's planetary orbital period is 1.58 Earth days, and the exomoon's orbital period is 0.231 Earth days. The exomoon transits the planet 7 times over a full planetary orbit. Figure~\ref{fig:GJ-lower}c shows the occultation scenarios and Fig.~\ref{fig:GJ-lower}d shows the transit scenarios. 
Each exomoon transit lasts about 81 minutes, as shown by the labelled dips in Fig.~\ref{fig:GJ-lower}a, while the planetary transit lasts about 60 minutes. Because GJ 1214 b's orbital period is much shorter than TRAPPIST-1 e's orbital period, the transit duration in Fig.~\ref{fig:GJ-lower}a appears visually much longer than those in Fig.~\ref{fig:trappist-lower}a, but the actual transit duration is comparable. 

The grey shades in Figs.~\ref{fig:GJ-lower}a and \ref{fig:GJ-lower}b show the 3$\sigma$ uncertainty of the observation data of GJ 1214 b's thermal phase curve. The 3$\sigma$ uncertainty is about 50 ppm \citep{Kempton_2023}, comparable to the exomoon's thermal phase curve amplitude. If an exomoon does exist around GJ 1214 b, extracting the exomoon signal through just one observation is not easy, unless we can capture the periodic signals of the exomoon-exoplanet transits. 
Unlike planet-star transits, which occur at a fixed orbital phase of the planet, the exomoon-exoplanet transits can occur at varying planetary orbital phases from one orbit to the next, unless the planet's orbital period is an exact integer multiple of the exomoon's orbital period. Averaging multiple thermal phase curves is likely to smooth out exomoon–exoplanet transit signals.

\subsection{Influence of surface thermodynamic parameters and rotation period on exomoon thermal phase curves\label{subsec:cp}}

\begin{figure*}
    \includegraphics[width=17cm]{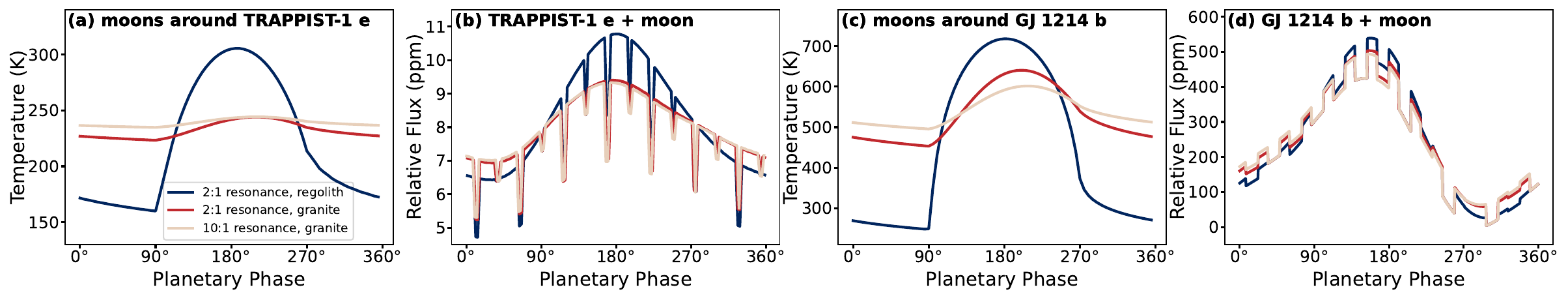}
    \centering
    \caption{Surface temperature and thermal phase curves of exomoons with different surface thermodynamic parameters and rotation periods. Panel (a): Meridional mean effective emission temperatures of 3000 km airless exomoons around TRAPPIST-1 e with: a 2:1 spin-orbit resonance and lunar regolith surface (blue line), a 2:1 spin-orbit resonance and granite surface (red line), and a 10:1 spin-orbit resonance and granite surface (pink line). Panel (b): Total thermal phase curves of TRAPPIST-1 e plus the according exomoons in panel (a). Panels (c) and (d): Same as panels (a) and (b) but for 6000 km airless exomoons orbiting GJ 1214 b.}
    \label{fig:cp}
\end{figure*}

Sections~\ref{TRAPPIST} and \ref{GJ} show that the exomoon's thermal phase curve amplitude depends on the radius and the day--night temperature difference of the moon. The influence of the exomoon's radius is clear and straightforward, so in this section we focus on the day--night temperature contrast and the factors that influence the contrast. In the simplified surface heat conduction model described in Sect.~\ref{subsec:SST}, from the subsurface energy equation (Eq.~\ref{equation_Tsoil}) and the boundary condition (Eq.~\ref{equation_Tboundary}), the analytic expression of day--night temperature difference can be obtained:
\begin{equation}
|A| = \frac{\left(1-\alpha\right) S}{b} \frac{1}{1+\sqrt{\omega\tau_D}\frac{1-i}{\sqrt{2}}},
\label{equation_deltaT}
\end{equation}
where $|A|$ is the day--night temperature difference, $\alpha$ is the exomoon albedo, $S$ is the stellar insolation, $\tau_D=(\rho c_p)^2 D/b^2$ is the mixed layer thermal relaxation timescale, $\rho c_p$ is the surface heat capacity, $D=\kappa_T/(\rho c_p)$ is the diffusivity, $b = 4 \sigma T^3$, and $\omega$ is the synodic angular velocity of the exomoon \cite[Chapter~7.4.2]{pierrehumbert2010principles}.

Obtaining the analytic expression of Eq.~\ref{equation_deltaT} needs perturbation method, which assumes that the solar flux only varies a little around a mean value \citep[Chapter~7.4]{pierrehumbert2010principles}. The diurnal solar flux is neither a sine function nor a small fluctuation around an average value. Therefore, Eq.~\ref{equation_deltaT} can only serve as a rough scaling of day--night temperature difference.

Based on Eq.~\ref{equation_deltaT}, five factors influence the day--night temperature difference: $\alpha$, $S$, $b$, $\omega$, and $\tau_D$. The albedo $\alpha$ has linearized influence on the day--night temperature difference. The parameter $b = 4 \sigma T^3$ essentially depends on the stellar insolation $S$. $S$ varies between different systems, and we have studied different systems in Sects.~\ref{TRAPPIST} and \ref{GJ}. Neither the rotation period nor the soil condition of exomoons are well known, so in what follows we focus on two factors: the synodic angular velocity ($\omega$) and the mixed layer thermal relaxation timescale ($\tau_D$), the latter of which essentially depends on the heat capacity ($\rho c_p$) and the thermal conductivity ($\kappa_T$). 

To study the influence of $\rho c_p$ and $\kappa_T$, we tried two types of surface condition. One set of thermal parameters is the same as lunar regolith in Sect.~\ref{TRAPPIST} ($\rho c_p = 10^{6}$ J m$^{-3}$ K$^{-1}$, and $\kappa_T = 0.01 \times 10^{-6}$ W m$^{-1}$ K$^{-1}$). The other set of thermal parameters is the same as granite, with heat capacity $\rho c_p = 2.02\times10^{6}$ J m$^{-3}$ K$^{-1}$, and thermal conductivity $\kappa_T = 2.9$ W m$^{-1}$ K$^{-1}$ \citep{pierrehumbert2010principles,Miranda_2018}. In general, granite surface has a smaller day--night temperature difference than regolith (red line versus blue line in Figs.~\ref{fig:cp}a and \ref{fig:cp}c), because both $\rho c_p$ and $\kappa_T$ are in the denominator of Eq.~\ref{equation_deltaT}. The larger $\rho c_p$ and $\kappa_T$ are, the slower the surface temperature reacts to the stellar flux variation, and the smaller the day--night temperature difference is.

The synodic angular velocity is given by $\omega=\omega_{orb}+\omega_{rot}$, where $\omega_{orb}$ is the angular velocity of the exomoon orbiting the planet, and $\omega_{rot}$ is the sidereal angular velocity of the exomoon. To simplify the problem, we fixed $\omega_{orb}$ and tested two types of spin-orbit resonance: 2:1 and 10:1. For exomoons around TRAPPIST-1 e, the orbital period is 0.876 Earth days. The rotation period is 0.438 Earth days for a 2:1 resonance and 0.0876 Earth days for a 10:1 resonance. For exomoons around GJ 1214 b, the orbital period is 0.231 Earth days, and the rotation period is 0.115 Earth days or 0.0231 Earth days. 
In general, fast-rotating exomoons have a smaller day--night temperature difference than slow-rotating exomoons (pink versus red lines in Figs.~\ref{fig:cp}a and \ref{fig:cp}c) because $\omega$ is in the denominator of Eq.~\ref{equation_deltaT}. The faster the exomoon rotates, the more evenly the exomoon is heated over the globe, and the smaller the day--night temperature difference is.

Not only does the day--night temperature difference vary between cases in Figs.~\ref{fig:cp}a and \ref{fig:cp}c, the hottest point also shows a phase lag in relation to each case. Both heat capacity ($\rho c_p$) and synodic angular velocity ($\omega$) influence the phase lag. Because $\rho c_p$ and $\kappa_T$ are higher for granite than for regolith, the mixed layer thermal relaxation timescale ($\tau_D$) for granite surface is about 693 times the $\tau_D$ for lunar regolith. For 2:1 spin-orbit resonance exomoons with lunar regolith surface, $\omega\tau_D \ll 1$, Eq.~\ref{equation_deltaT} simplified to $|A| = {\left(1-\alpha\right) S}/{b}$. The surface temperature almost reaches instantaneous equilibrium and the phase lag is negligible. For exomoons with granite surface, $\omega\tau_D > 1$, the imaginary part in Eq.~\ref{equation_deltaT} cannot be neglected anymore, and leads to phase lag. The faster the exomoon rotates, the larger $\omega\tau_D$ is, and the larger the phase lag is, as shown by the pink and red lines in Fig.~\ref{fig:cp}a and \ref{fig:cp}c. When $\omega\tau_D \gg 1$, Eq.~\ref{equation_deltaT} simplifies to $|A| = {\left(1-\alpha\right) S}/{b} / (\sqrt{\omega\tau_D}\frac{1-i}{\sqrt{2}})$. The coefficients of the real part and the imaginary part are equal, and the phase lag reaches the upper limit, $\pi/4$. 

The number of eclipses depends only on the period of the exoplanet orbiting the star and the period of the exomoon orbiting the exoplanet. Varying the surface or the rotation period of the exomoon does not influence the time or the duration of the eclipses, but influences the depth of exomoon-exoplanet eclipses, as shown by Figs.~\ref{fig:cp}b and ~\ref{fig:cp}d.

\begin{figure*}[htbp]
    \sidecaption
    \includegraphics[width=12cm]{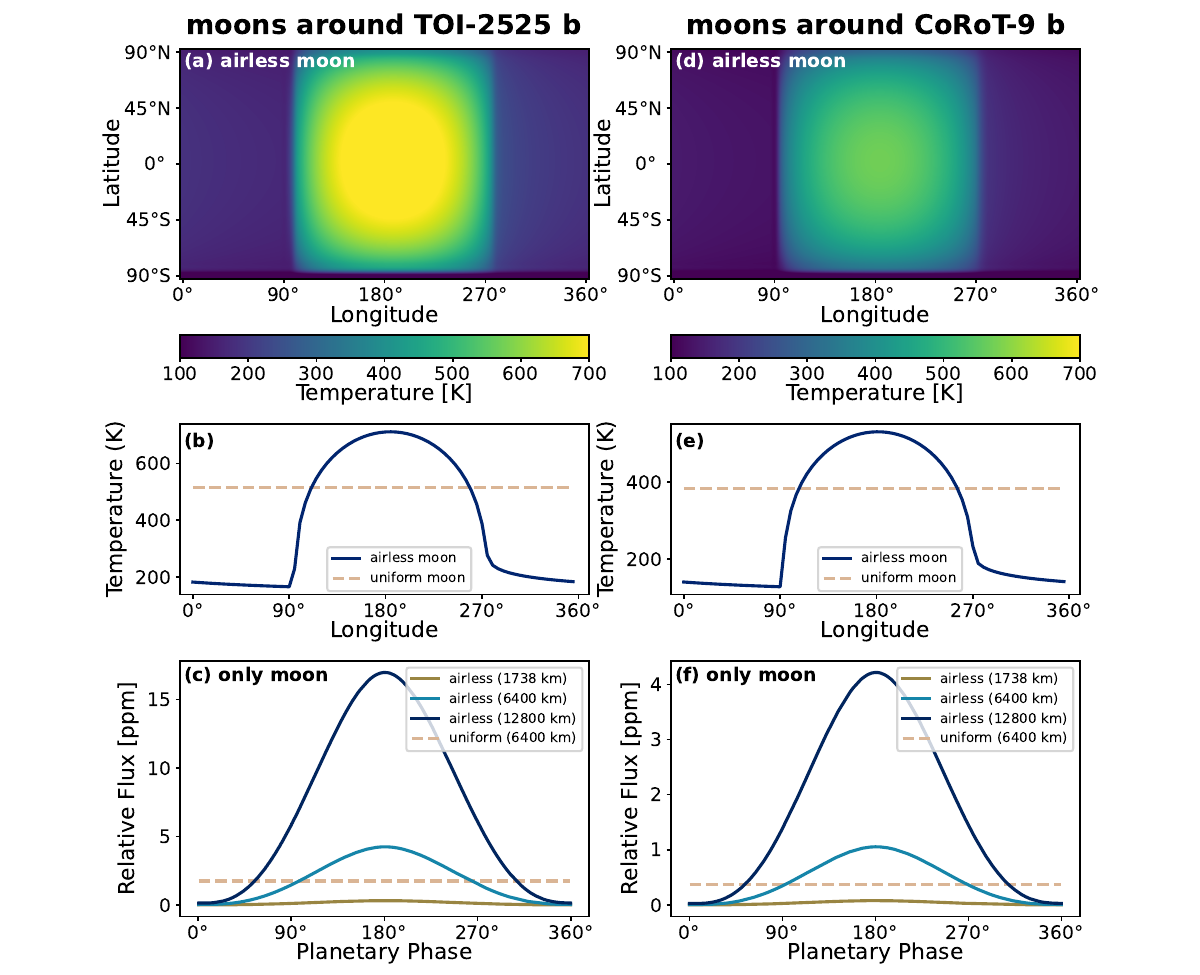}
    \caption{Surface temperature and thermal phase curves for hypothetical exomoons in two longer-period planetary systems. Left column: Results for the TOI-2525 b system (planetary orbital period of 23.3 days). Panel (a): Snapshot of the surface temperature distribution for a tidally locked airless exomoon. Panel (b): Meridional mean effective emission temperature for an airless exomoon (blue line) and a uniform-temperature exomoon (dashed line). Panel (c): Thermal phase curves of the exomoons. The dashed line shows the thermal phase curve of a uniform-temperature exomoon with a radius of 6400 km. The exomoons orbit at 40\% Hill sphere (285338 km) around TOI-2525 b, with an orbital period of 3.4 Earth days. Right column: Same as the left column but for the CoRoT-9 b system (planetary orbital period 95.3 days). The exomoons orbit at 1570880 km around CoRoT-9 b, and the orbital period is 13.9 Earth days.}
    \label{fig:2in1}
\end{figure*}

\subsection{Exomoons around longer-period exoplanets \label{subsec:long_period}}

\begin{figure}
    \includegraphics[width=\linewidth]{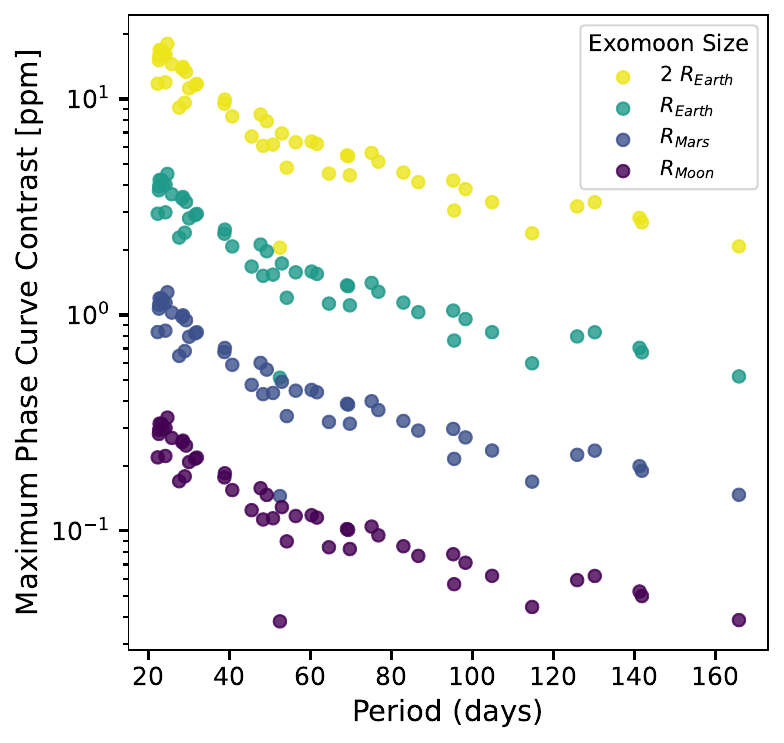}
    \centering
    \caption{Thermal phase-curve amplitude survey for hypothetical airless exomoons as a function of the host planet’s orbital period. Colours indicate different assumed exomoon radii. The x-axis shows the orbital periods of the host planet, and the y-axis shows the exomoon thermal phase curve amplitude. All hypothetical exomoons orbit at 40\% of the Hill radius around their host planets.}
    \label{fig:survey}
\end{figure}

\begin{figure*}
    \includegraphics[width=1\textwidth]{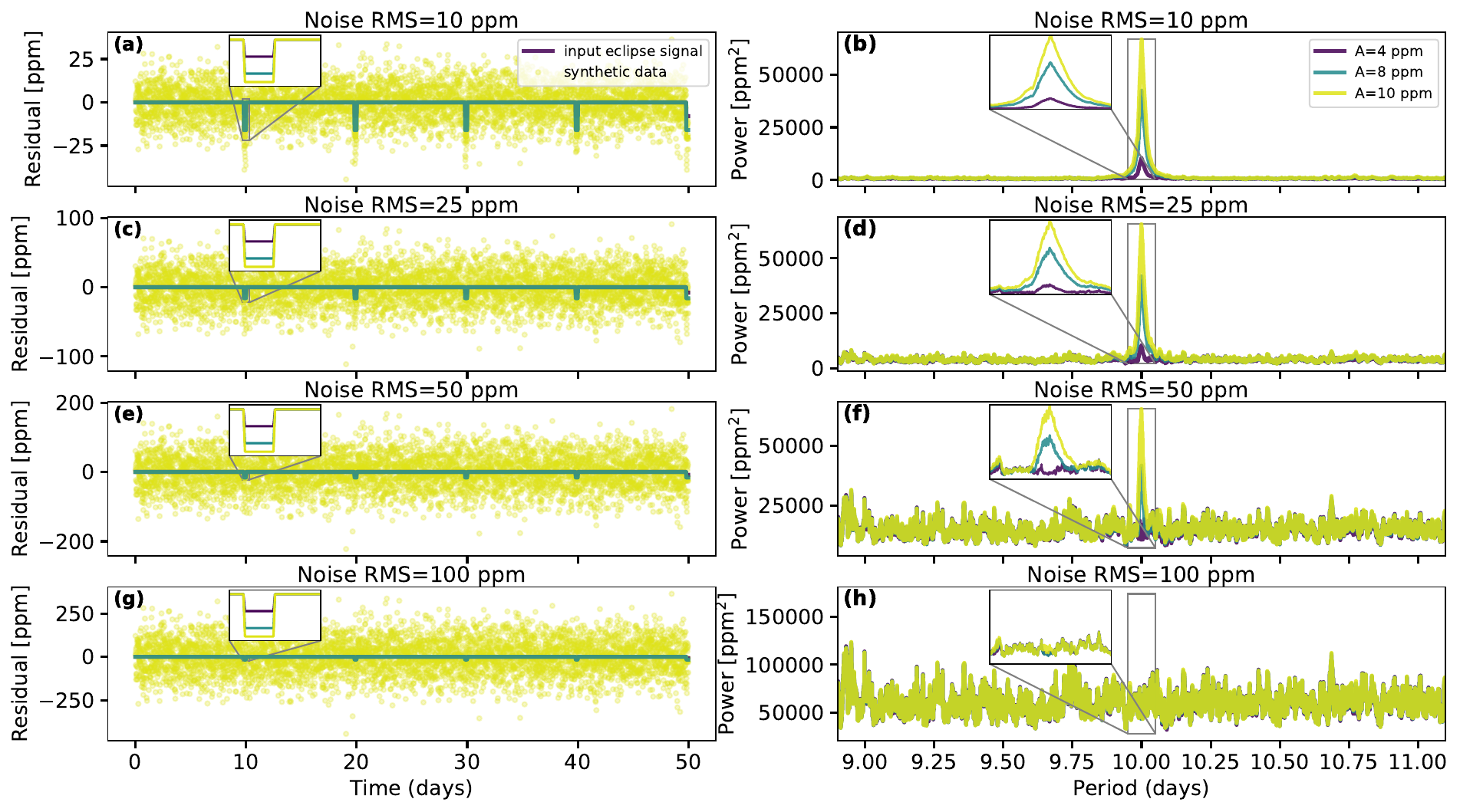}
    \centering
    \caption{Synthetic exomoon-exoplanet mutual eclipse signals (left column) and their according periodograms (right column). Each row corresponds to a different white noise level, with RMS values ranging from 10 to 100 ppm. The dots in the left column shows the injected signals combined with noise. Different colours represent different exomoon thermal phase curve amplitudes (A), from 2 ppm to 10 ppm. The solid lines in the right column show the corresponding periodograms. Each hypothetical signal has a sampling frequency of 15 minutes and a total observation time of 180 days.}
    \label{fig:fap}
\end{figure*}

Sections~\ref{TRAPPIST} and \ref{GJ} study hypothetical exomoons around short-period exomoon planets. Although exomoons around short-period planets are not strictly forbidden by orbital dynamics 
\citep{Domingos_2006,Cassidy_2009,Kisare_2023}, several previous studies suggest that the long-term stability of such systems is limited \citep{Barnes_2002,Sasaki_2012,Sucerquia_2019,Dobos_2021}. In this section we therefore extend our analysis to exomoons orbiting longer-period exoplanets, in order to assess the applicability and limitations of our method in more astrophysically realistic regimes.

The left column in Fig.~\ref{fig:2in1} shows the surface temperature and thermal phase curves of exomoons orbiting at 40\% of the Hill radius around TOI-2525 b. The figure layout follows that of Fig.~\ref{fig:GJ-upper}, but for a host planet with a longer orbital period. TOI-2525 b has an orbital period of 23.3 days, and an Earth-sized airless exomoon produces a thermal phase curve amplitude of approximately 5 ppm. For CoRoT-9 b, which has an orbital period of 95.3 days, an Earth-sized airless exomoon yields a thermal phase curve amplitude of only $\sim$1 ppm, as shown by the right column of Fig.~\ref{fig:2in1}. 
At fixed stellar parameters, the longer the planetary orbital period is, the smaller the exomoon day--night temperature contrast is, and the smaller the exomoon thermal phase curve amplitude is.

We analysed a selected subset of confirmed transiting exoplanets with orbital periods longer than 20 days, focusing on systems that yield relatively large exomoon thermal phase curve amplitudes in order to illustrate upper-limit signal strengths. For each system, we considered hypothetical airless exomoons orbiting at 40\% of the Hill radius, and computed the  thermal phase curve amplitudes for a range of exomoon radii. The results are shown in Fig.~\ref{fig:survey}. For large exomoons (up to two Earth radii), the thermal phase curve amplitude can reach up to $\sim$20 ppm. Overall, the phase curve amplitude exhibits a clear negative correlation with the orbital period of the host planet.

In real observations involving exomoon-exoplanet mutual eclipses, the combined thermal phase curve appears as a periodic signal, as shown in Fig.~\ref{fig:GJ-lower}a. After subtracting the planetary transit model from the data, the periodic signal contributed by the exomoon is buried in the residuals. Is it possible to extract the underlying signal from noisy residuals? We performed an idealized injection-recovery test using synthetic exomoon--exoplanet mutual eclipse signals embedded in white noise.

The solid lines in the left column of Fig.~\ref{fig:fap} show the input exomoon-exoplanet mutual eclipse models. Each synthetic dataset spans a total observation time of 180 days, with a 4.8-hour mutual eclipse occurring every 10 days. Different colours indicate exomoon thermal phase curve amplitudes ranging from 2 to 10 ppm. The yellow dots show the simulated observations sampled every 15 minutes embedded in white noise. The root mean square (RMS) of the white noise increases from 10 ppm to 100 ppm in Figs.~\ref{fig:fap}a--\ref{fig:fap}g. The corresponding periodograms are shown in the right column.

For a white-noise level of 50 ppm RMS, periodic signals with amplitudes around 10 ppm can be identified using the box-fitting least squares method \citep{Kov_cs_2002}, as shown in Fig.~\ref{fig:fap}f. The signal peak at 10 days indicates the mutual eclipse period. For future telescopes, if the noise level can reduced to an RMS of 25 ppm, detecting exomoons through thermal phase curves becomes more feasible (Fig.~\ref{fig:fap}d). 

Compared with conventional exomoon–star transit searches, our method has two key advantages: substantially reduced requirement on the total observational time span and the predictable, periodic nature of the signal.
Transit-based exomoon detection generally relies on repeated observations of multiple planet–moon–star transits. Even for large exomoons ($\sim$1.5 Earth radii), 5–6 such events are typically required to achieve a reliable detection \citep{simon2015cheops}, with significantly more observations needed for smaller satellites. For long-period exoplanets, the time it takes to accumulate the required number of transits may span several planetary orbits, or even several years. Moreover, exomoon–star transits occur at different orbital phases in each observation, complicating the data analysis.
In contrast, our method focuses on exomoon–exoplanet mutual eclipses, which are periodic and can occur multiple times within a single planetary orbit. As a result, multiple exomoon signatures can in principle be captured within a much shorter overall observation span, making this approach a more time-efficient complement to transit-based exomoon searches.

\section{Discussion and conclusions} \label{sec:discussions}

\subsection{Conclusions}
Our main conclusions are as follows:
\begin{itemize}
\item Airless exomoons can significantly increase the total thermal phase curve amplitudes of the planet–moon systems due to large day--night temperature contrasts.
\item For exomoons smaller than $\sim2\,R_{e}$ orbiting planets with periods longer than 20 days, the exomoon thermal phase curve amplitudes are typically below $\sim$20 ppm.
\item Mutual eclipses between the planet and the exomoon lead to a periodic feature that is potentially crucial for detecting exomoons.
\item Ignoring the thermal contribution of an exomoon can lead to an overestimation of the planetary day--night temperature contrast or the planetary effective emission temperature.
\end{itemize}

\subsection{Discussion}
We now discuss the implications of our results. Airless exomoons can maintain high day--night temperature contrasts, increasing the total thermal phase curve amplitudes of exoplanet--exomoon systems. A survey of hypothetical exomoons around a wide range of exoplanets shows that for an exomoon with a radius of less than 2 Earth radii and whose host planet's orbital period is longer than 20 days, the thermal phase curve amplitude is less than 20 ppm. Large exomoons around close-in planets can, in principle, produce large thermal phase curve amplitudes and have substantial day--night temperature differences. Such configurations are dynamically extreme and serve as upper-limit demonstrations of the signal strength. For example, an almost-Earth-sized airless exomoon around GJ 1214 b produces a $\sim$100 ppm thermal phase curve amplitude.

If observing the thermal phase curve of a planet--moon system, and retrieving the planetary temperature distribution without extracting the moon signal, the planetary day--night temperature contrast will be significantly overestimated. For exomoons with uniform temperatures, the day--night contrast of its host planet is not strongly affected, but the inferred planetary effective emission temperature may be overestimated, as shown by the dashed lines in Figs.~\ref{fig:trappist-lower}b and \ref{fig:GJ-lower}b.

Detecting mutual transit signals between the exomoon and its host planet is crucial to separating the lunar and planetary thermal phase curves.  Mutual eclipses produce periodic, dent-like features in the combined thermal phase curves. Using idealized simulations, we show that such signals may be identifiable under favourable noise conditions, although our simplified model does not account for variations in eclipse depths or long-term trends in the planetary thermal phase curve. Different period-search techniques, such as the box-fitting least squares and Lomb–Scargle methods, detect these faint periodic features equally well. A more dedicated detection tool is left for future work.

Due to the Kozai mechanism and tidal evolution, an exomoon is likely to be coplanar with the host planet's equatorial plane \citep{Porter_2011}. Therefore, if a transiting exoplanet has a low obliquity, the exomoon is likely to transit its host planet. 
In our Solar System, the orbital planes of the majority of large satellites are indeed closely aligned with their host planet's equatorial planes. For example, the orbital inclinations of Phobos and Deimos are roughly 1$^{\circ}$ relative to the Martian equatorial plane \citep{murchie2015phobos}; the orbital inclinations of the Galilean satellites range from 0.04$^{\circ}$ to 0.51$^{\circ}$ \citep{bills2000galilean}; and the orbital inclinations of Enceladus and Titan are less than 0.4 $^{\circ}$ \citep{tholen2000planets}. 
In contrast, some small and distant satellites do have large orbital inclinations. For instance, the Jovian `Himalia' satellite family has orbital inclinations ranging from 25$^{\circ}$ to 55$^{\circ}$, but their radii are less than 90 km \citep{holt2018cladistical,tholen2000planets}. 

When calculating the exomoon's surface temperature distribution, we considered only the stellar incident flux, and did not consider the influence of thermal emission from the planet or the stellar flux reflected by the planet on the exomoon's surface temperature. Given the bond albedo $\alpha=0.3$, the average planetary emission received by the moon, $F_{e} = \left(1-\alpha\right)S R_{planet}^2 / 4 a_{moon}^2$, plus the average planetary reflection received by the moon, $F_{r} < \alpha S R_{planet}^2 /4 a_{moon}^2$, is less than 0.6\% of stellar flux ($S$) for exomoons around TRAPPIST-1 e, and less than 6\% of $S$ for exomoons around GJ-1214 b. This is consistent with the results of \cite{Hinkel_2013}, suggesting that the thermal and reflected radiation from the planet do not make a significant contribution to the exomoon's flux range. 

Tidal forces are crucial to the orbital stability of exomoons, but the intensity of tidal dissipation remains poorly constrained. In a simplified theory, considering the tidal dissipation caused by the planet, the evolution of the exomoon's semi-major axis could be written as 
\begin{equation}
\dot{a}_{moon} = sign(\Omega_p-n) \frac{3 k_p}{Q_p} \frac{M_{moon}}{M_{planet}} \left( \frac{R_{planet}}{a_{moon}} \right)^5 n a_{moon},
\label{equation_tide}
\end{equation}
where $k_p$, $Q_p$, and $\Omega_p$ are the tidal Love number, the quality factor, and the rotation angular velocity of the exoplanet, and $n$ and $a_{moon}$ are the mean motion and semi-major axis of the exomoon \citep{murray1999solar,zhou2024yarkovsky}. The tidal love numbers, $k_p$, are roughly determined, usually with a value of around 0.3 \citep{murray1999solar,Saliby_2023}, but the quality factor, $Q_p$, is poorly constrained for exoplanets. Setting $k_p=0.3$, and assuming the exomoon's initial semi-major axis is 0.4 Hill radii, the stability time of exomoons varies a lot with $Q_p$, as shown in Fig.~\ref{fig:dadt}a. 

Substituting the planetary parameters in Eq.~\ref{equation_tide} with the stellar parameters, we find that the tidal dissipation caused by the star is at least 10$^{30}$ orders smaller than the tidal dissipation caused by the planet. We therefore neglected the influence from the star in our subsequent estimations.
\begin{figure}
    \includegraphics[width=\linewidth]{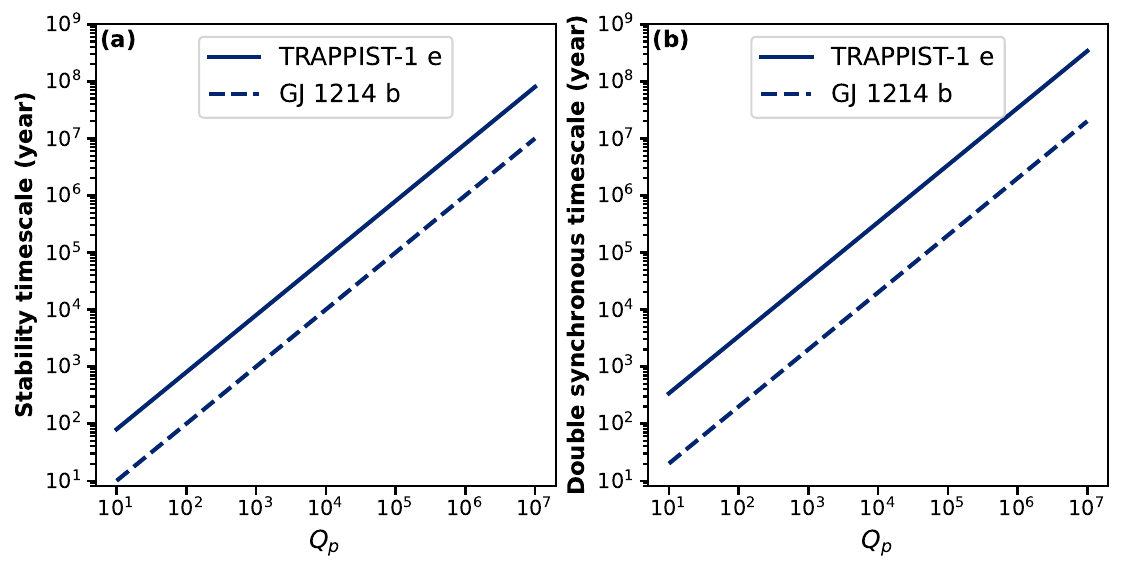}
    \centering
    \caption{Stability time of hypothetical exomoons around TRAPPIST-1 e and GJ 1214 b (a) and the timescale for an exomoon and exoplanet to reach a double synchronous state (b), under different planetary quality factor ($Q_p$) values.}
    \label{fig:dadt}
\end{figure}

For exomoons around TRAPPIST-1 e, the stability timescale is of the order of a thousand years if $Q_p \sim $ 10$^2$, but if $Q_p \sim $ 10$^5$, the stability timescale can reach the order of a million years. For rocky planets in our Solar System, $Q_p$ ranges from 10 to 10$^2$ \citep{murray1999solar}. However, simulations find that many factors influence $Q_p$, such as mantel viscosity, planetary size, orbital eccentricity, the transient response duration in the primary creep, and tidal frequencies \citep{efroimsky2009tidal,Clausen_2015,Xiao_2020}. If the mantle viscosity of TRAPPIST-1 e reaches 10$^{22}$ Pa s, $Q_p$ could reach 10$^5$ \citep{Xiao_2020,Saliby_2023}. For GJ 1214 b, a sub-Neptune, $Q_p$ is estimated to reach 10$^4$ to 10$^5$ \citep{james2024thermal}, and the stability timescale for its exomoons is of the order of 10$^5$ years. Figure~\ref{fig:dadt}a shows that the stability timescale ranges from hundreds to millions of years, so we cannot completely rule out the possibility of observing exomoons in these systems.   

Furthermore, tidal forces influence not only the semi-major axis of the exomoon, but also the rotation period of the exoplanet, given by \cite{murray1999solar},
\begin{equation}
\dot{\Omega}_{p} = -sign(\Omega_p-n) \frac{3 k_p}{2\beta_p Q_p} \frac{M_{moon}^2}{M_{planet}(M_{planet}+M_{moon})} \left( \frac{R_{planet}}{a_{moon}} \right)^3 n^2 ,
\label{equation_omega}
\end{equation}
where $\beta_p$ is a parameter of moment of inertia ($I_p = \beta_p M_{planet}{R_{planet}}^2$), which is set to Earth's value of 0.33 \citep{yoder1995astrometric} in our calculations. With tidal dissipation transferring the angular momentum between the spin of the planet and the orbit of the moon, the exomoon and the exoplanet can reach a double synchronous state, in which the spin periods of both the planet and the moon are equal to the period of the moon orbiting the planet. In other words, the same side of the exomoon is always facing the same side of the exoplanet. The timescale for the exomoon and the exoplanet to achieve a double synchronous state is about the same as the tidal decay timescale of the exomoon's semi-major axis, as shown in Fig.~\ref{fig:dadt}b. More detailed studies have found this scenario possible, especially for massive moons, which are more likely to survive around inner exoplanets, preventing inner exoplanets from spiralling into their host stars \citep{makarov2023pathways,efroimsky2024synchronous}. In future studies, we plan to carry out more detailed estimations using dynamical models with tidal dissipation processes.

The Yarkovsky effect of the exomoon, which is the radiation force raised on the afternoon side caused by the phase lag of the hottest point, is also negligible for exomoons. Using Eq. 13 in \cite{brovz2006yarkovsky}, we find that the Yarkovsky acceleration of the exomoon's semi-major axis is of the order of 10$^{-9}$ AU My$^{-1}$. The acceleration is negligible because the mass of the exomoon is too high compared to the typical metre- to kilometre-sized asteroids that are affected by the Yarkovsky effect \citep{vokrouhlicky1998diurnal,vokrouhlicky1999complete,bottke2006yarkovsky,Vokrouhlick__2015,zhou2024yarkovsky}.

The diversity and complexity of exoplanetary systems continue to expand the range of possible environments in which exomoons may exist. Exploring this broader parameter space can help identify promising targets and optimize exomoon search strategies. However, those `unlikely' stable or short-lived moons may still produce observable signatures. If such exomoons are detected, they will help us refine our orbital theories. With future telescopes of higher precision, the method proposed in this study can be applied to observe exomoons, opening up a wider possibility of exomoon detection.

\begin{acknowledgements}
      We thank Su Wang, Lixiang Gu, Daniel Koll, Douglas N. C. Lin, and  Wenhan Zhou for their helpful discussions. J.Y. acknowledges support from the National Natural Science Foundation of China (NSFC) under Grant 42550102.
\end{acknowledgements}

\bibliography{aa55379-25}
\bibliographystyle{aa}

\end{document}